# Variation of Fractal Properties of Two-Component Systems in Their Evolution Process

K.S. Baktybekov, S.G. Karstina*, E.N. Vertyagina, A.A. Baratova

*Eurasian national university named after L.N. Gumilyov, 5 Munaitpasov st., Astana, 010008, Kazakhstan, tel./fax: (3172) 35-67-83, e-mail: baktybek@emu.kz*
*\* Karaganda state university named after E.A. Buketov, 28 Universitetskaya st., Karaganda,100028, Kazakhstan*

**Analysis of experimental kinetic dependences and results of the computer modeling of pairwise interactions in the heterogeneous system have shown that a character of its fractal property variation in the evolution process would be different at distinguished temperature ranges and dependent on the proportion of reagent concentrations and their initial distribution type.**

## 1. Introduction

In open non-equilibrium indigested systems as a result of transfer processes of energy, matter or information the formation of new structures is possible, in which fractal dimension depends on their generation rate, process mechanisms working in the systems being studied, and also on some intrinsic parameters, which characterize the state of a system as a whole. The forming spatial and temporal structures can be described within the bounds of the model of a system dynamic self-organization, which is accompanied by a decrease of entropy production. On the other hand a transition from unstable structures to more stable ones can be considered as a non-equilibrium phase transition. Self-organization and non-equilibrium phase transitions lead to non-classical dependences, which determine system state as a whole at its intrinsic and external parameter variations. For example, the formation of localized structures with high concentration of reagents and accordingly high efficiency of current transfer processes leads to non-classical behavior of luminescence decay in a system of excited molecules and their nearest surrounding. In this connection at the analysis of experimental kinetic dependences it is necessary to take into account the features of the initial reagent distribution, the change of their distribution type as a result of photo-physic processes, and therefore the temporal dependence of appropriate reaction rate constants, which lead to the variation of the fractal properties of the formed clusters.

## 2. Results and discussion

Modification of a reagent distribution type on a surface as a result of annihilation processes can be simulated by the probabilistic cellular automata method. For this reason, the computer modeling of two-component system evolution as a result of pairwise interactions has been performed at the temperature range of 153-273 K at initial multifractal, monofractal and chaotic distributions. The proportions of reagent concentrations were 1:1, 1:2, 1:4, 1:10 and 1:40. The simulation was made on 2d-lattice, which have had 500*500 nodes. The radius of different kind particle interaction was limited by the minimal distance between the lattice nodes. The relative mobility of reagents was determined depending on the temperature of the process. The particle interaction in the system was set randomly, the probability of the pairwise interaction was selected as equal to 1/5, ½ and 1. The values of parameters, which characterize the system state at the moment, such as information entropy, conditional interaction potential, Rényi generalized fractal dimensions and multifractal spectrum function were calculated by the multifractal analysis method.

The model in use reproduces the evolution of the system, in which hetero-annihilation interactions are possible in donor-acceptor pairs. Molecular adsorbates of bengal pink dye (BP), which were donors of the triplet energy, and molecular adsorbates of anthracene aromatic hydrocarbon (An), which were acceptors of the triplet energy, were selected as the system. The observed kinetics of the annihilation delayed fluorescence has the non-exponential character at the temperature range of 173 to 273 K at the proportions of reagent surface density of 1:1, 1:4, and 1:40. The the nonuniformity parameters, the quenching sphere radius and the transfer rate constant, which were calculated on the basis of experimental curves, have a complex dependence on the matrix temperature and the relative proportion of reagent concentration [1, 2]. The results of the computer modeling of the system evolution with the initial multifractal distribution give a close explanation of the observed experimental dependencies. At different moments of observation period over the system state there was performed multifractal analysis (MFA) for the obtained reagents distributions in the simulated system. MFA allowed the estimation of such parameters, which characterize a structural state of a system, as fractal dimensions and information entropy.

In the general case the generalized fractal dimensions $D_q$ [3] hold information about thermodynamic conditions of a system formation. Value $D_0$ $(q = 0)$ is a Hausdorff fractal dimension; it determines the highest possible value of a system fractal dimension. Information entropy $S_{sp}$ [4] is a measure of information, which a structure obtains at the transition from one level of organization to another. The amount of information which is necessary for determination of the state of each structure elements can be defined from the specific value of information entropy.

Temperature dependences $D_0$ и $S_{sp}$ obtained from MFA results at the 100$^{th}$ iteration of kinetic curves allow marking out three distinctive sections (fig. 1).

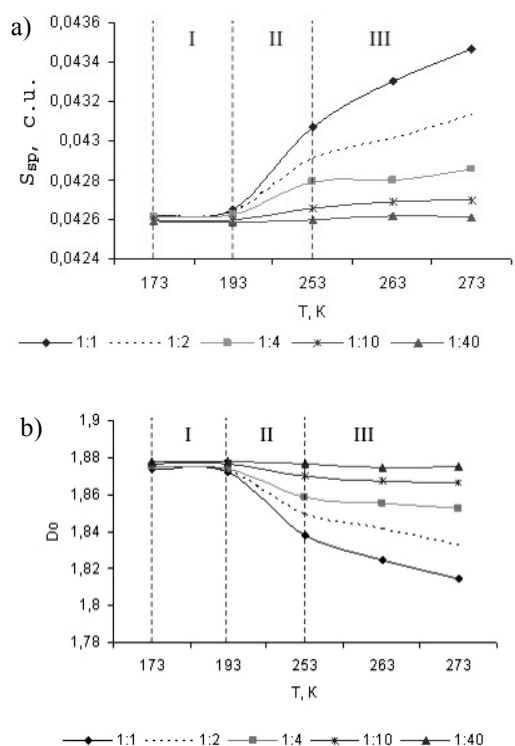

Fig. 1. Dependences of the values of information entropy (a) and fractal dimension (b) on the matrix temperature at the different proportions of reagent molecule concentrations.

At the temperature range of 173-193 K (section 1) the fractal dimension $D_0$ of the system and the value of information entropy $S_{sp}$ remain constant for all proportions of reagent concentrations. This fact can be explained by the formation of stable, spatially limited structures and local interaction between the particles which form the system at these temperatures. At the same time the probability of interaction between the particles, which belong to the structure, does not influence the kinetics of the processes and therefore the fractality of the structures (fig. 2a).

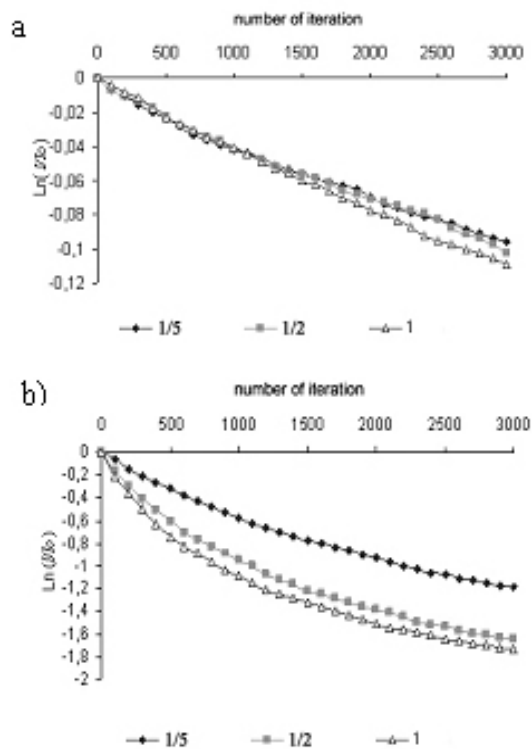

Fig. 2. Kinetic curves obtained as a result of computer modeling at different probabilities of two kind particle interactions: a) 183 K; b) 273 K.

The second section, which corresponds to temperatures 193-253 K, shows that the rates of changing of parameters $D_0$ и $S_{sp}$ vary essentially depending upon the proportion of particle number of each kind. At this temperature range the formed clusters begin diffusing and strongly interacting with each other due to the diffusion of particles. Reconfiguration of the structure as a whole takes place in the system; it appears as a decrease of the system fractal dimension (fig. 1b). It is obvious that their interaction efficiency is the highest at equal concentrations of reagents. Thus, at the increase of matrix temperature the variation of the fractal dimension will occur faster; and the structures being formed will be less stable than those at the first temperature range. At the same time the increase of interaction probability leads to the growth of information entropy $S_{sp}$., The less the amount of information is, which one gets about the system, the higher the rate of $S_{sp}$ changing is. It is important to note that despite the fact that $S_{sp}$ and $D_0$ parameters were calculated independently of one another, one can see their clear correlation in fig. 1. The number of interaction acts per time unit decreases at the concentration increase of one kind of particles. Also, the destruction of the system slows down, which is partly compensated by the reconfiguration of the structure, and this has an impact on $S_{sp}$ and $D_0$ parameters behavior.

At higher temperatures of the matrix there arises a combined macrocluster instead of local microclusters. It leads to the slowing down of the rate of particle interaction per time unit, and this has an influence on the delay of variation of system fractal parameters (section III in fig. 1). The kinetics of the processes is determined by the probability of interaction and, hence, by the dimension of the clusters being formed (fig. 2b) at this temperature range. At the same time a great increase of concentration of one kind of particles in comparison with the concentration of another kind of particles leads to such reagent distribution where $D_0$ and $S_{sp}$ parameters become constant at all studied temperature ranges. Pairwise interactions become ineffective and the system can be considered as pseudo-monomolecular.

## 3. Conclusion

The peculiarities of values variation of fractal dimension and information entropy at the evolution of a heterogeneous system, which have been considered in the paper, can be explained by a supposition that the proportion of interacting particles concentrations in the system and the matrix temperature are control parameters; they have a significant influence on photo-processes in the system being studied. In the area of low temperatures the system tends to keep its intrinsic multifractal parameters irrespective of the donor or acceptor concentration. The increase of the matrix temperature leads to the loss of the system stability and the replacement of mechanism of electron interaction energy transfer; the evolution process of the system is determined by the proportion of interacting particle concentrations and their interaction probability.